\documentclass[journal,twoside]{IEEEtran}

\usepackage{amssymb}
\usepackage{latexsym}
\usepackage{amsmath,amssymb,amsfonts}
\usepackage{algorithmic}
\usepackage{graphicx}
\usepackage{textcomp}
\usepackage{booktabs}
\usepackage{colortbl}
\usepackage{tabularx}
\usepackage{multirow}
\usepackage{xcolor}
\usepackage{hyperref}
\usepackage{makecell}

\definecolor{mygray}{gray}{0.9}
\usepackage{colortbl}
\hypersetup{
    colorlinks=true,
    linkcolor=green,
    filecolor=magenta,      
    urlcolor=cyan,
    citecolor=green
}

\def\BibTeX{{\rm B\kern-.05em{\sc i\kern-.025em b}\kern-.08em
    T\kern-.1667em\lower.7ex\hbox{E}\kern-.125emX}}
\begin{document}
\title{MediViSTA: Medical Video Segmentation\\via Temporal Fusion SAM Adaptation\\for Echocardiography}

\author{Sekeun Kim, Pengfei Jin, Cheng Chen, Kyungsang Kim, Zhiliang Lyu, Hui Ren, \\
Sunghwan Kim, Zhengliang Liu, Aoxiao Zhong, Tianming Liu, Xiang Li, Quanzheng Li 
\thanks{Sekeun Kim,  Pengfei Jin, Cheng Chen, Kyungsang Kim, Zhiliang Lyu, Hui Ren, Xiang Li, and Quanzheng Li are with the Center of Advanced Medical Computing and Analysis, Massachusetts General Hospital and Harvard Medical School, Boston, MA 02114, USA}
\thanks{Sunghwan Kim is with the Center of Advanced Medical Computing and Analysis, Massachusetts General Hospital and Harvard Medical School, Boston, MA 02114, USA and Department of Thoracic and Cardiovascular Surgery, Gyeongsang National University Changwon Hospital, Gyeongsang National University College of Medicine, Changwon, Republic of Korea} 
\thanks{Aoxiao Zhong is with Harvard John A. Paulson School of Engineering and Applied Sciences, Harvard University, Cambridge, MA 02138, USA}
\thanks{Zhengliang Liu and Tianming Liu are with the School of Computing, The University of Georgia, Athens, GA 30602, USA}
}

\maketitle
\begin{abstract}
Despite achieving impressive results in general-purpose semantic segmentation with strong generalization on natural images, the Segment Anything Model (SAM) has shown less precision and stability in medical image segmentation. In particular, the original SAM architecture is designed for 2D natural images and is therefore not support to handle three-dimensional information, which is particularly important for medical imaging modalities that are often volumetric or video data. In this paper, we introduce MediViSTA, a parameter-efficient fine-tuning method designed to adapt the vision foundation model for medical video, with a specific focus on echocardiographic segmentation. To achieve spatial adaptation, we propose a frequency feature fusion technique that injects spatial frequency information from a CNN branch. For temporal adaptation, we integrate temporal adapters within the transformer blocks of the image encoder. Using a fine-tuning strategy, only a small subset of pre-trained parameters is updated, allowing efficient adaptation to echocardiographic data. The effectiveness of our method has been comprehensively evaluated on three datasets, comprising two public datasets and one multi-center in-house dataset. Our method consistently outperforms various state-of-the-art approaches without using any prompts. Furthermore, our model exhibits strong generalization capabilities on unseen datasets, surpassing the second-best approach by 2.15\% in Dice and 0.09 in temporal consistency. The results demonstrate the potential of MediViSTA to significantly advance echocardiographical video segmentation, offering improved accuracy and robustness in cardiac assessment applications.
\end{abstract}


\section{Introduction}
\label{sec:introduction}
\IEEEPARstart{C}{ardiovascular} disease remains the leading cause of mortality worldwide \cite{di2024heart}, emphasizing the critical importance of accurate and timely cardiac assessments. Echocardiography, as the first-line imaging modality for cardiac evaluation, provides 2D images with dynamic information of the heart. This non-invasive technique allows for detailed assessments from various probe orientations, offering crucial insights into cardiac health. However, interpreting echocardiograms presents significant challenges. It requires manual segmentation by experienced sonographers, with the accuracy of the assessment largely dependent on their ability to integrate information from both preceding and subsequent frames to fully capture cardiac dynamics. This reliance on individual expertise can introduce variability in interpretations, which can affect patient outcomes \cite{leclerc2019deep}. To address these challenges, cardiac segmentation has emerged as a fundamental tool in quantifying key metrics such as ejection fraction and ventricular volume, which are essential for diagnosis and treatment planning \cite{klaeboe2019echocardiographic}. Deep learning techniques have shown promise in automating and improving the accuracy of cardiac segmentation \cite{leclerc2020lu, ouyang2020video}. However, achieving satisfactory results remains difficult due to the inherent limitations of ultrasound image quality and the scarcity of annotated data.

Recently, the progress of foundation models trained on large-scale and diverse datasets has shifted the paradigm of model design. These models have gained significant attention across various domains due to their ability to generalize across diverse tasks, reducing the need for task-specific models and enabling broad applicability. The Segment Anything Model (SAM) \cite{hu2304skinsam} is a visual foundation model designed for prompt-based image segmentation, trained on an extensive dataset containing more than 1 billion masks in 11 million natural images. With its large-scale training and flexible architecture, SAM has exhibited remarkable zero-shot performance on a variety of tasks. This naturally raises the question of whether SAM can be applied effectively to medical imaging, potentially addressing the ongoing challenge of limited annotated datasets in this field. However, due to the substantial domain gap between natural and medical images, recent evaluations of SAM on medical imaging tasks have revealed that its zero-shot capabilities, with or without prompts, are insufficient for direct application to medical images.

These differences require domain-specific adaptations to take advantage of the transfer ability of the original SAM for medical image segmentation. The recent consensus to adapt SAM for medical applications is rapidly growing. Common approaches to adapting SAM can be divided into two groups: interactive segmentation with SAM’s prompt design and automatic segmentation with adapter or parameter-efficient transfer learning (PETL). In the context of interactive SAM adaptation \cite{wu2023medical, deng2023sam}, these approaches aim to enhance the support for more comprehensive and user-friendly prompts by leveraging the SAM prompt design during fine-tuning. However, providing the appropriate prompts for medical images is a complex task. The adaptation of SAM for automated medical image segmentation, including methods such as adapters \cite{ma2023segment, bui2023sam3d} and parameter-efficient transfer learning (PETL) techniques such as Low-Rank Adaptation (LoRA) \cite{hu2021lora} and FacT \cite{jie2023fact}, utilizes low-rank representations to reduce the number of trainable parameters, showing promising results \cite{wu2023medical, zhang2023customized, zhu2024sam}. However, these methods focus primarily on 2D adaptation, overlooking the critical temporal information present in medical video data, which is essential to capture dynamic changes and improve segmentation accuracy.

\begin{figure}
    \centering
    \includegraphics[width=1.0\linewidth]{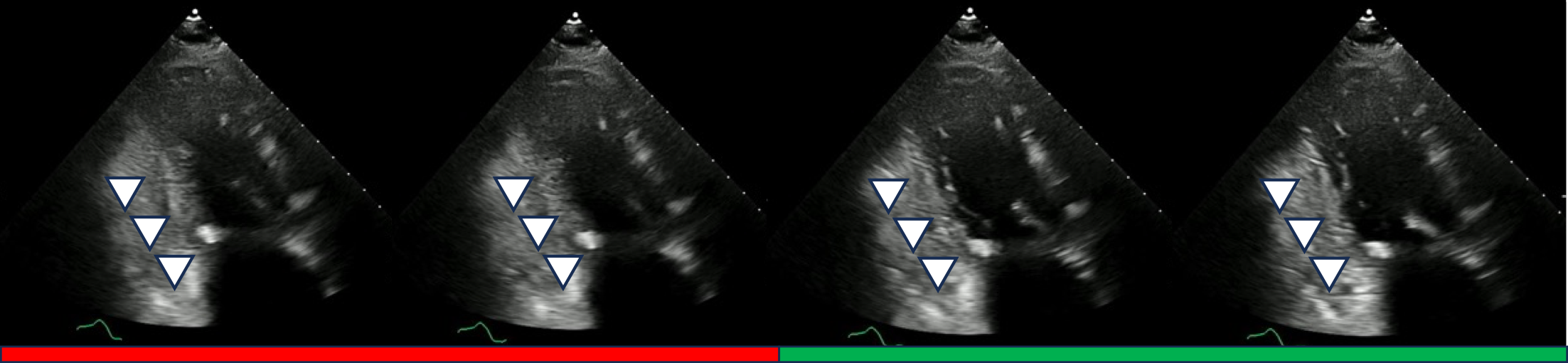}
    \caption{The dynamic nature of echocardiography: cardiac boundaries often appear indistinct or poorly defined in earlier frames  \textcolor{red}{red}, becoming clearer in later frames  \textcolor{green}{green}. Limitations of memory-based methods include their reliance on processing only past and current frames, which prevents them from utilizing information from future frames.}
    \label{figure0}
\end{figure}

Segmenting medical videos, especially echocardiography, presents distinct challenges because of the intricate spatial and temporal properties of moving organs. For example, the low signal-to-noise ratio often makes it difficult to accurately delineate cardiac boundaries. Furthermore, the intricate three-dimensional motion of the heart \cite{lee2018three}, with anatomical structures appearing intermittently and disappearing across frames, further complicates accurate segmentation. These challenges require advanced techniques that can capture both the spatial details of individual frames and the temporal continuity of cardiac motion across sequences. Recent efforts, MEMSAM \cite{deng2024memsam}, have extended SAM for echocardiography segmentation by incorporating a memory network \cite{oh2019video}, but rely exclusively on past frames, which can result in performance limitations, particularly when dealing with echocardiography datasets where future frames contain valuable predictive information, as in Fig. \ref{figure0}. To address these challenges, it is essential to develop video segmentation approaches that effectively capture spatial features within individual frames while leveraging information from both past and future frames.

In this paper, we propose MediViSTA, a parameter-efficient method to adapt SAM from 2D to 2D+T for medical video segmentation. MediViSTA effectively leverages spatio-temporal features to address the low signal-to-noise ratio and dynamic characteristics of echocardiography. We introduce a frequency feature fusion module which utilizes CNN branch and integrate features into SAM's transformer with cross-attention mechanism. Our approach employs the PETL technique, FacT \cite{jie2023fact}, which uses tensorization-decomposition to improve fine-tuning efficiency. This approach mitigates most of the pretrained weights and updates lightweight parameters to minimize modifications while enhancing performance. To bridge the gap between 2D natural image and medical video data, we introduce a temporal fusion attention to use adjacent frames for model robustness. For the mask decoder, we utilize the intermediate features of the transformer to connect the feature to the corresponding stages of the mask decoder via skip connections, thus restoring the original prediction resolution. We perform extensive evaluations, comparing our method with domain-specific and SOTA SAM adaptation approaches. The results show that our method significantly outperforms existing techniques, particularly excelling in zero-shot analysis on unseen echocardiography datas. Our major contributions are summarized as follows:


\begin{itemize}
\item {We propose a parameter-efficient fine-tuning method, called MediViSTA, to adapt SAM for medical video input. This approach makes most of the pretrained weights reusable while achieving strong performance in medical video segmentation.
}
\item {We introduce spatio-temporal adapters consisting of frequency feature fusion and a temporal-fusion attention mechanism, which effectively capture multi-scale frequency spatial features and dynamic temporal patterns in echocardiography.}

\item{We conducted extensive experiments on two public datasets as well as multi-center curated dataset. The results show that our MediViSTA significantly outperforms the performance of SOTA methods, especially in a challenging, unseen multicenter dataset.}
\end{itemize}

\begin{figure*}[h]
    \centering
    \includegraphics[width=1.0\linewidth]{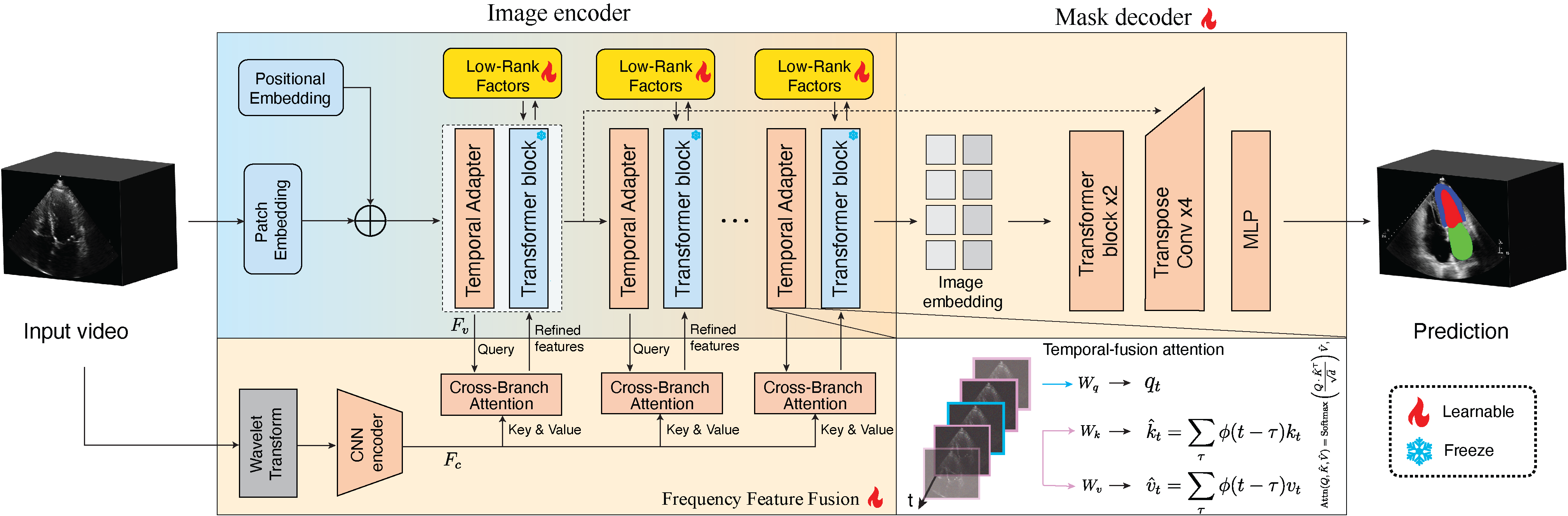}
    \caption{Overview of the proposed methods. The frequency feature fusion module integrates spatial frequency features from CNN branch and ViT’s global features through cross-attention, enriching the information available for segmenting the target. The temporal adapter extends SAM’s pretrained 2D backbone to incorporate third-dimensional information through a temporal-fusion attention mechanism. This mechanism captures visual cues by integrating information from surrounding frames, enabling more accurate analysis of each ultrasound frame within its temporal context. The pretrained SAM backbone is fine-tuned using a parameter-efficient transfer learning approach with Factor-tuning (FacT). Additionally, a multiscale mask decoder is employed to recover the original input resolution for segmentation. }
     \label{method}
\end{figure*}
    
\section{Related Work}
\subsection{Foundation Models in Computer Vision}
Foundation models pre-trained in large-scale data have shown exceptional generalization in a wide range of computer vision tasks. Florence \cite{yuan2021florence} focuses on learning universal visual-language representations, making it adaptable to tasks such as visual question answering and video retrieval. SAM \cite{kirillov2023segment}, trained on the SA-1B dataset with 11 million natural images, stands out for its effectiveness and broad applicability, especially in zero-shot image segmentation. Although SegGPT \cite{wang2023seggpt} and SEEM \cite{zou2023segment} are also designed for general image segmentation, they are pretrained on smaller datasets compared to SAM. Recently, SAM-2 \cite{ravi2024sam} was introduced to extend the capabilities of SAM, enabling it to handle more complex scenarios in video analysis. This extension allows SAM-2 to process temporal image sequences, making it well-suited for tasks that require an understanding of spatial continuity across frames. However, despite the success of these foundation models in computer vision, their performance often declines when applied to medical downstream tasks \cite{roy2023zero, yan2024biomedical}.

\subsection{SAM Adaptation in Medical Imaging}
To address this performance gap between the natural image and the medical image, many studies have explored ways to leverage the transferability of SAM for medical imaging segmentation \cite{ma2023segment, zhang2023customized, bui2023sam3d, wu2023medical}. 
 Med-SAM \cite{ma2023segment} maintains the original SAM network structure but uses more suitable bounding box prompts for the medical domain and focuses on fine-tuning the mask decoder. SAMed \cite{zhang2023customized} uses a LoRA technique to fine-tune the image encoder. For 3D medical data, SAM3D \cite{bui2023sam3d} merged adjacent slices into the batch dimension for SAM's 2D image encoder and proposed the 3D decoder to incorporate depth information. Modification of SAM's image encoder to adapt medical images such as Med-SA \cite{wu2023medical}, which introduces a medical adapter that transposes the spatial dimension of the input embeddings to the depth dimension. This approach processes different dimensional information of inputs and addresses the disparity between 2D images and 3D volumetric medical images such as MRI and CT scans. MedSAM-2 \cite{zhu2024medical} adapts SAM-2 by taking advantage of its memory bank for 2D and 3D segmentation tasks. The aforementioned models are primarily designed and evaluated on 3D volumetric datasets, and their application to video datasets remains largely unexplored.
 
 Domain-specific adaptations for ultrasound imaging have been developed, including SAM-US \cite{lin2023samus}, SAM-Att \cite{zhu2024sam}, and MEMSAM \cite{deng2024memsam}. SAM-US \cite{lin2023samus} employs dual image encoders with CNN and ViT branches. The image embeddings from these branches are then carefully merged using cross-attention. SAM-Att \cite{zhu2024sam} fine-tuned SAM's pretrained image encoder using a LoRA strategy. In addition, it incorporates a convolutional block attention module in the decoder to enhance performance on echocardiography images. Although SAM-US \cite{lin2023samus} and SAM-Att \cite{zhu2024sam} have shown promising results, these methods are limited to processing 2D images, not fully utilizing temporal information. Space-time memory networks (STM), such as MEMSAM \cite{deng2024memsam}, propagate semantic information along the temporal dimension. These models require a good initialization, relying on previous temporal frames for effective segmentation.
 
\section{Methodology}
In this section, we will introduce how we adapt the original 2D SAM architecture for medical video segmentation. We will first present an overview of the original SAM structure. We will then provide a detailed explanation of each component in our method, including temporal information adaptation, frequency feature fusion using a CNN branch, and parameter-efficient fine-tuning of the pretrained SAM block. Additionally, we have modified the mask decoder by incorporating progressive up-sampling to restore the original input resolution. An overview of our framework is illustrated in Fig. \ref{method}.

\subsection{Overview of SAM}
We provide an overview of the SAM \cite{kirillov2023segment}, which consists of three core components: an image encoder, a prompt encoder, and a mask decoder. The image encoder utilizes the structure of the Vision Transformer (ViT)  \cite{dosovitskiy2020image}, which has been pretrained using a masked autoencoder approach in a self-supervised manner. The prompt encoder encodes various prompts such as points and boxes into embedding representations. The mask decoder comprises a self-attention block and bidirectional cross-attention blocks. After the attention blocks, the feature map is up-sampled, and segmentation masks are generated by fully connected layers. The authors have officially released three pretrained models: ViT-B/16, ViT-L/16, and ViT-H/16, corresponding to the base, large, and huge models, respectively, in terms of model size.

\subsection{Temporal Fusion Attention}
To adapt the 2D SAM for 2D + T medical video input, we propose a temporal fusion adaptation within the image encoder, integrating a temporal adapter into SAM's 2D transformer blocks. As illustrated in Figure \ref{method}, our model processes video inputs represented as a tensor $X \in \mathbb{R}^{B \times C \times T \times H \times W}$, where $B$ denotes the batch size, $C$ represents the number of input channels, $T$ indicates the total number of frames, and $H$ and $W$ correspond to the height and width of each frame, respectively.

Recently, numerous studies have integrated cross-attention mechanisms into various architectures, primarily to boost model performance in tasks that involve complex relational data spanning different modalities or datasets. Specifically, cross-attention allows tokens $x_i$ from one sequence to utilize the key and value vectors of another sequence $\hat{x}_i$.
The fundamental goal of incorporating cross-attention is to improve a model's capacity to attend to and synthesize information from diverse sources more effectively, thereby increasing its ability to comprehend and process interrelated, yet distinct, data streams \cite{kumar2016ask,tan2019lxmert}.

Motivated by the unique challenges associated with echocardiography, including a low signal-to-noise ratio and intermittent visibility or obscurity of certain anatomical structures across different frames of the echocardiographic sequence, we sought to develop an analytical method that integrates temporal information throughout the entire sequence. Consequently, we propose a temporal-fusion attention mechanism.

Specifically, rather than employing query, key, and value vectors generated from a single frame as in traditional self-attention mechanisms, we propose to utilize key and value vectors generated from multiple frames to perform cross-attention. Given the inherent continuity in echocardiographic sequences, we employ a weighted average of key and value vectors for cross attention, where the weights are dependent on the temporal differences:
\begin{align}
q_t = W^{Q}x_{t}, \quad k_t = W^{K}x_{t}, \quad v_t = W^{V}x_{t}, 
\label{crosstime1}
\end{align}
\begin{align}
\hat{k}_t = \sum_{\tau} \phi(t,\tau) k_{\tau} , \quad \hat{v}_t = \sum_{\tau} \phi(t,\tau) v_{\tau}, 
\label{crosstime2}
\end{align}\
\begin{align}
\text{Attn}(Q, \hat{K}, \hat{V}) = \text{Softmax}\left(\frac{Q \cdot \hat{K}^\top}{\sqrt{d}} \right)\hat{V}, \label{att2}
\end{align}
where $Q=[q_1,\dots,q_n],\hat{K}=[\hat{k}_1,\dots,\hat{k}_n],\hat{V}=[\hat{v}_1,\dots,\hat{v}_n]$, and $\phi$ is a kernel function designed to modulate the influence of frames based on their temporal distance. For this kernel function, we use a Gaussian kernel. 
\begin{align}
\phi(t, \tau) = \exp \left( -\frac{(t - \tau)^2}{2 \sigma^2} \right),
\end{align}
\( \tau \) represents the position of the query frame, and the value of \( \sigma \) controls the spread of the Gaussian curve. This Gaussian kernel \( \phi \) modulates the influence based on the temporal distance, with \( \sigma \) determining how much the neighboring frames affect each other. A smaller \( \sigma \) concentrates the influence around the query frame, while a larger \( \sigma \) allows for broader temporal influence across more frames. In our ablation studies, we will demonstrate that this temporal-fusion attention mechanism, by incorporating information across multiple frames, significantly improves the performance in segmentation.

\subsection{Frequency Feature Fusion Module}
The SAM's image encoder reduces the input image resolution by a factor of 16 using patch embeddings, which may be insufficient to capture fine-grained spatial details essential for segmentation tasks. Furthermore, while the basic blocks of SAM consist of transformers, which are powerful in capturing global characteristics through self-attention, they lack the ability to effectively represent local features. In this regard, we introduce a frequency feature fusion module (FFM) consisting of CNN blocks. This module takes as input the wavelet-transformed image, which is decomposed into its respective frequency sub-bands, enabling multi-scale analysis. The proposed method implements a discrete wavelet transform (DWT) using the Haar wavelet basis \cite{heil1989continuous}. Specifically, the image is decomposed into four sub-bands: approximation coefficients (LL) and detailed coefficients for horizontal (LH), vertical (HL), and diagonal (HH) components. The CNN branch extract frequency features then we inject to original SAM's transformer blocks. To align CNN and ViT features, cross-branch attention is applied to fuse the feature maps. Specifically, for the feature maps $\mathcal{F}_v$ from the ViT branch and $\mathcal{F}_c$ from the CNN branch. The cross-branch attention treats the global context from the ViT as the query (Q) and the local details from the CNN as the keys (K) and values (V). This fusion mechanism allows the model to enhance global representations with localized details, improving its overall performance on segmentation.

\subsection{Parameter-efficient Fine-tuning}
To effectively adapt the pretrained SAM for the medical domain, we employ a SOTA parameter-efficient fine-tuning strategy, Factorized Tuning (FacT) \cite{jie2023fact}. Instead of fully fine-tuning all layers of the SAM image encoder, we utilize FacT, a technique that leverages rank redundancy in transformer models. FacT approximates dense weight updates during fine-tuning using low-rank factors, shared across layers for greater efficiency. This low-rank adaptation technique reduces the number of trainable parameters while preserving most of the pretrained knowledge. FacT assumes that the dense weight increments required for fine-tuning can be approximated by low-rank factors. Specifically, the weight increases \(\Delta W\) are decomposed into three components: \(U \in \mathbb{R}^{d \times r}\), \(V \in \mathbb{R}^{d \times r}\), and \(\Sigma \in \mathbb{R}^{r \times r}\), where \(d\) denotes the dimension of the feature and \(r \ll d\) is the rank of the decomposition. The factors \(U\) and \(V\) are shared across all layers, while \(\Sigma\) is unique to each layer. The weight increments can then be expressed as:


\begin{align}
\Delta W_{i,j,k} &= \sum_{t_{1}=1}^{r} \sum_{t_{2}=1}^{r} \Sigma_{i,t_{1},t_{2}} U_{j,t_{1}} V_{k,t_{2}}, \label{eq:1}
\end{align}

where \( i \) indicates the index of the transformer layer, and \( j, k \) are the indices of dimensions. This decomposition significantly reduces the number of trainable parameters, making fine-tuning more efficient. In our framework, these low-rank updates are applied specifically to the query and value transformations within each transformer block, ensuring that most of the pretrained SAM weights remain frozen. The modified query and value transformations after fine-tuning are given by:

\begin{align}
W_{q/v} = W_0 + U \Sigma_{q/v} V^T,
\end{align}

where \(W_{q/v}\) refers to the fine-tuned query or value weights, and \(W_0\) denotes the pretrained SAM weights. By employing this strategy, we extend the capabilities of SAM to handle complex medical data while preserving the generalizability of the model and minimizing computational overhead.

\subsection{Multi-scale Mask Decoder Adaptation}
The original SAM mask decoder, despite its efficient and lightweight design, faces challenges when applied to medical imaging tasks, particularly in producing high-resolution medical segmentation. The original mask decoder consists of two transformer layers, four transposed convolutional layers, and a single multilayer perceptron; this results in a substantial reduction in output resolution, producing segmentations that are only 1/4 of the input's resolution. This resolution reduction significantly limits the decoder's effectiveness in medical imaging tasks, where precise delineation of small anatomical structures or lesions is crucial. To address these challenges, progressive multi-scale upsampling strategies have proven particularly useful for medical image segmentation \cite{hatamizadeh2103transformers, tang2022self}. Specifically, we redesigned the SAM's mask decoder with four CNN layers that interact with intermediate image features from ViT. To achieve this, we uniformly divide the image encoder into four stages and progressively upsample the resolution, establishing connections between the feature maps at each stage and the decoder, while preserving the prompt encoder architecture. In our experiments, we have observed that progressive up-sampling in the decoder shows superior outcomes compared to approaches without these features, showing the efficacy and simplicity of our approach.

\begin{table*}
\caption{Segmentation performance of the proposed method with state-of-the-art methods on CAMUS, EchoNet-dynamic, and multi-center datasets. EchoNet-dynamic dataset only includes LV$_{endo}$ annotation. The Dice coefficient, Hausdorff distance, Average Symmetric Surface Distance, and temporal consistency, respectively. The hyphen indicates that the model failed to segment the region of interest, with a Dice score of less than 30\%. Best results are denoted as \textbf{bold}.}
	\centering
	\begin{tabularx}{\textwidth}{l*{8}{>{\centering\arraybackslash}X}}
        \toprule
        \multirow{3}{*}{\makecell{(a) CAMUS/ \\EchoNet-dynamic}} & \multicolumn{4}{c}{LV$_{endo}$} & \multicolumn{4}{c}{LV$_{epi}$} \\
        \cmidrule(r){2-5} \cmidrule(r){6-9}
        & Dice$\uparrow$ & dH(mm)$\downarrow$ & dA(mm)$\downarrow$ & \textit{L}$\downarrow$ & Dice$\uparrow$ & dH(mm)$\downarrow$ & dA(mm)$\downarrow$ & \textit{L}$\downarrow$ \\
        \midrule
        LUNet  \cite{leclerc2020lu} & 91.39/ 90.13 & 7.66/ 10.52 & 1.73/ 4.82   &0.08/ 1.51 & 80.04 & 11.10 & 1.84 & 0.22 \\
        EchoNet \cite{ouyang2020video} &92.63/ 91.10 & 6.23/ 8.54 & 1.19/ 3.55   & 0.07/ 1.10 & 84.18 & 8.42 & 1.21 & 0.14  \\
        \midrule
        Unet++ \cite{zhou2019unet++} & 92.65/ 89.17  & 6.23/ 8.10 & 1.19/ 4.31   & 0.07/ 0.61 & 83.41 & 9.44 & 1.42 & 0.12  \\
        U-Transformer  \cite{petit2021u} & 93.10/ 90.88  & 6.51/ 8.04 &  0.78/ 2.18 & 0.06/ 0.68 & 86.47 & 8.23 &1.03 & 0.11  \\
        STM  \cite{oh2019video} & 91.40/ 90.21  & 6.58/ 9.10&  1.70/ 5.10 & 0.14 /0.80 & 84.20 & 9.51 & 1.82 & 0.19 \\
        
        \midrule
        Med-SAM\cite{ma2023segment} & 91.03/ 89.21 & 6.80/ 8.63  & 1.76/ 4.02 & 0.28/ 0.69 & 86.08 & 12.58 & 2.43  & 0.73 \\
        Med-SA\cite{wu2023medical} & 93.22/ 90.30 & 5.10/ 7.61  & 1.44/ 3.21 & 0.09/ 0.52 & 88.61 & 9.27  & 1.21 & 0.12    \\ 
        SAM-US\cite{lin2023samus}  & 92.61/ 90.08 & 6.40/ 7.89  & 2.30/ 4.68 & 0.46/ 0.65 & 88.07 & 9.40  & 1.38  & 0.15  \\
        SAM-Att\cite{zhu2024sam}   & 88.60/ 87.01 & 11.40/ 7.58 & 5.68/ 6.84 & 0.31/ 0.89 & 84.50 & 12.11 & 3.45  & 0.32  \\
        SAM3D\cite{bui2023sam3d}   & 92.42/ 91.50 & 4.68/ 5.04  & 0.88/ 1.92 & 0.08/ 0.51 & 87.11 & 10.55 & 3.22  & 0.23  \\
        MEMSAM\cite{deng2024memsam}  & 92.35/ 91.03 & 6.40/ 5.30  & 2.29/ 3.45 & 0.29/ 0.61 & - & - & -  & -  \\
        MedSAM-2\cite{zhu2024medical}  & 91.38/ 91.43 & 6.18/ 5.08  & 2.10/ 3.24 & 0.27/ 0.59 & - & - & -  & -  \\        
        \midrule
        MediViSTA (BBox) & \textbf{94.62/ 93.80} & \textbf{4.52/ 3.80} & \textbf{0.60/ 1.28}  & \textbf{0.06/ 0.35} & - & - & - & -\\
        MediViSTA  & 94.42/ 93.05 & 4.54/ 4.10 & 0.68/ 1.40  & 0.05/ 0.30 & \textbf{90.24} & \textbf{8.30}  &\textbf{0.98}& \textbf{0.08}\\
        \bottomrule
        \bottomrule
    \end{tabularx}
   \begin{tabularx}{\textwidth}{lXXXX XXXX XXXX}
        \multirow{3}{*}{(b) Multi-center} & \multicolumn{4}{c}{LV$_{endo}$} & \multicolumn{4}{c}{LV$_{epi}$} & \multicolumn{4}{c}{LA} \\
        \cmidrule(r){2-5} \cmidrule(r){6-9} \cmidrule(r){10-13}
        & Dice$\uparrow$ & dH(mm)$\downarrow$ & dA(mm)$\downarrow$ & \textit{L}$\downarrow$ & Dice$\uparrow$ & dH(mm)$\downarrow$ & dA(mm)$\downarrow$ & \textit{L}$\downarrow$ & Dice$\uparrow$ & dH(mm)$\downarrow$ & dA(mm)$\downarrow$ & \textit{L}$\downarrow$ \\
		\midrule
        LUNet  \cite{leclerc2020lu} & 87.91  & 14.21   & 6.10  & 0.16 & 75.11 & 15.63  & 7.10 &  0.18 & 84.54   & 18.48  & 4.91  & 0.09 \\
        EchoNet \cite{ouyang2020video} & 88.12  &  13.30 & 5.44  & 0.10 & 74.20 & 13.95 & 6.90 & 0.13 & 85.70   & 16.22  & 4.88  & 0.06 \\
        \midrule
		Unet++  \cite{zhou2019unet++} & 85.12  & 16.34   & 6.42  & 0.19 & 72.22 & 19.43  & 7.49 &  0.21 & 83.52   & 23.04  & 5.91  & 0.07 \\  
        U-Transformer  \cite{petit2021u} & 88.20  & 12.92 & 4.18  & 0.20 & 80.98 & 11.34 & 5.45 & 0.14 & 87.44 &  13.52 & 3.92 &  0.08  \\
        STM  \cite{oh2019video} & 87.30  & 13.48 & 6.28  & 0.48 & 79.80 & 13.22 & 7.35 & 0.22 & 85.80 & 15.08 & 4.82 &  0.19 \\
        \midrule
        Med-SAM\cite{ma2023segment} & 89.08  & 14.75 & 5.86 & 0.39 & 81.60  & 13.05 & 6.45  & 0.27  & 87.65 & 15.24 & 5.16 & 0.24 \\
        Med-SA\cite{wu2023medical} & 89.19 & 13.05 & 5.02 & 0.21  & 82.16  & 12.52  & 6.15 & 0.17 & 89.00 & 10.16 & 3.88  & 0.17\\ 
        SAM-US\cite{lin2023samus} & 88.45  & 13.75 & 5.92 & 0.25 & 78.67  & 14.10 & 6.77  & 0.24  &  87.72 & 14.19 & 4.10 & 0.22 \\
        SAM-Att\cite{zhu2024sam} & 88.42  & 16.21 & 7.30 & 0.26 & 81.22 & 12.09 & 7.02  & 0.26  &  86.87 & 16.20 & 6.05 & 0.35  \\
        SAM3D\cite{bui2023sam3d} & 90.60  & 10.40 & 5.68 & 0.21  & 82.11  & 12.22 & 4.53  & 0.12 &  89.91 & 9.01 & 4.10 & 0.14 \\
        MEMSAM\cite{deng2024memsam} & 89.96  & 11.58 & 4.22 & 0.25 & -  & - & -  & - &  86.64 & 12.05 & 4.18 & 0.27 \\
        MedSAM-2\cite{zhu2024medical} & 90.03  & 11.10 & 4.82 & 0.24 & -  & - & -  & - &  87.48 & 11.85 & 4.11 & 0.26 \\
        \midrule
        MediViSTA (BBox) & \textbf{92.59} & \textbf{8.10} & \textbf{2.10} & \textbf{0.05} & - & - & - & - & \textbf{92.05} & \textbf{7.68} & \textbf{2.71}  & \textbf{0.05}\\ 
        MediViSTA &  92.05 & 8.38 & 2.15  & 0.05 & \textbf{85.80} & \textbf{7.46} & \textbf{3.82} & \textbf{0.08} & 91.24 & 7.95 & 2.80 & 0.06 \\
		\toprule
	\end{tabularx}
\label{table1}
\end{table*}

\section{Experiments}
We trained our model on the CAMUS dataset and evaluated it on three datasets: CAMUS, EchoNet-Dynamic, and a multi-center internal dataset. A key motivation for employing pretrained foundation models is their strong potential to generalize effectively to unseen data. To evaluate this capability, we employ a specific experimental strategy: training the model solely on the CAMUS dataset \cite{leclerc2019deep} and evaluating it not only on the CAMUS dataset but also on the EchoNet-Dynamic dataset \cite{ouyang2020video} and our multi-center internal dataset. This approach allows for a comprehensive evaluation of the model’s capacity to generalize in diverse data distributions and clinical settings.

\subsection{Dataset for fine-tuning and evaluation}
The CAMUS dataset comprises 2D echocardiography of 1,000 patients, including apical two-chamber (A2CH) and four-chamber (A4CH) views for 500 of these patients. For training purposes, the data set provides sparse annotations throughout the cardiac cycle, specifically in end-diastole (ED) and end-systole (ES) frames from 402 patients. For evaluation, three datasets are used: CAMUS, EchoNet-Dynamic, and a multi-center internal dataset. The CAMUS test data set consists of 98 patient A4CH view images with LV$_{endo}$, LV$_{epi}$ annotations over time \cite{painchaud2022echocardiography}. The data set includes a mixture of normal patients and those requiring medical attention, and patients with an ejection fraction lower than 45\% considered at pathological risk. The EchoNet-Dynamic \cite{ouyang2020video} contains 10,030 A4CH view videos. This dataset only provides the LV$_{endo}$ in the ED and ES phases. For the multicenter data set, we curated a multicenter data set comprising B-mode echocardiography from 100 patients, including A2CH and A4CH. Data were acquired from patients receiving care at Massachusetts General Hospital and Brigham and Women's Hospital between 2017 and 2022. Imaging data were obtained using two types of scanners from different vendors: GE and Philips. Each manufacturer contributed equally, providing samples from 50 patients each, ensuring a balanced distribution in the study.

Two experienced clinicians labeled and reviewed the delineation process. The annotations include the boundaries of the LV$_{endo}$, LV$_{epi}$, and LA at the ED and ES phases. Due to intermittent noise and image obscuration, clinicians meticulously examined adjacent frames in video sequences to pinpoint and define accurate boundaries, following the recommendations of the American Society of Echocardiography \cite{lang2015recommendations}. This annotation process was performed using Slicer 3D software \cite{fedorov20123d}, a tool well regarded for its precision in medical imaging analysis. This study is exempted from human subjects, as determined by the Mass General Brigham Institutional Review Board protocol 2021P000681.

\begin{figure*}
\centering
\includegraphics[width=1.0\linewidth]{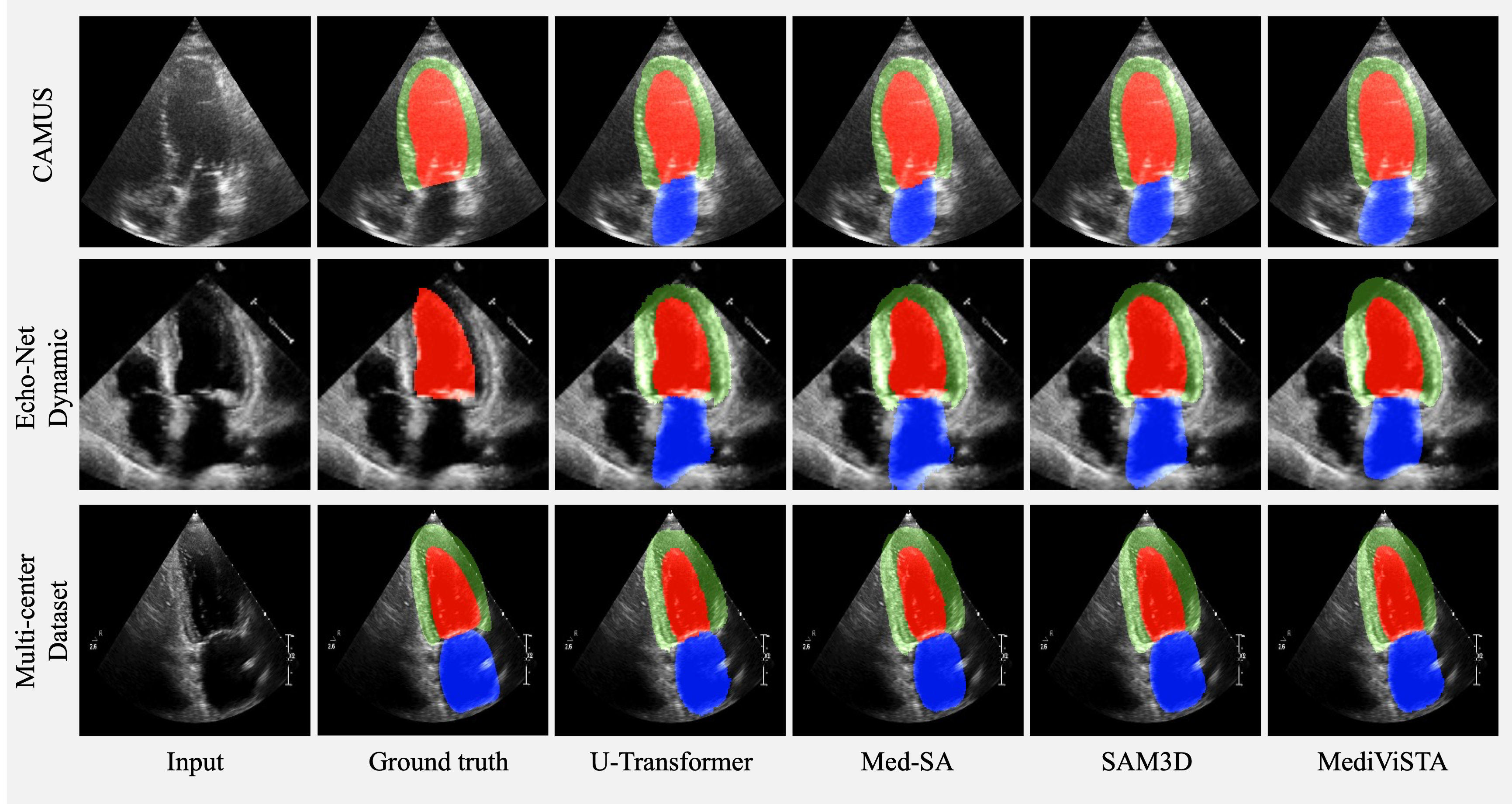}
\caption{Visual comparison of segmentation results on CAMUS, EchoNet-dynamic, and multi-center dataset. Best-performing methods are represented, including one from non-SAM methods and two from SAM-based methods. The \textcolor{blue}{blue}, \textcolor{red}{red}, and \textcolor{green}{green} regions denote the LA, LV$_{endo}$, and LV$_{epi}$, respectively.}
\label{overlay}
\end{figure*}

\subsection{Implementation details}
For preprocessing, we followed the same procedure as in \cite{kim2024assessment} to remove patient information and other irrelevant data. The image intensities were normalized to a scale between [0, 1] using min-max normalization. In order to standardize the input dimensions, we sampled frames between the ED and ES phases to have at least one cardiac cycle. To ensure consistent frame counts across all samples, we implemented an additional sampling strategy. 
 If the initial sampling interval between the ED and ES frames, including annotated masks, is less than one cardiac cycle (empirically set to 32 frames), we retrieve more frames after one cycle. We set the base learning rate to 1e-4 and used the AdamW optimizer for model training. To enhance the robustness and generalization of the model, we implemented data augmentation techniques during the training process. These included random flipping, scaling, and contrast adjustments of the input images. All experiments presented in this paper were conducted using PyTorch, taking advantage of the computational resources of 8 NVIDIA A100 GPUs.
 
The annotation of echocardiographic videos poses a significant challenge due to the sparse distribution of labels across frames. To address this partially labeled data scenario, we selectively applied the loss function only to frames with available class labels, effectively masking out unlabeled frames during the backpropagation process. Then, our network parameters are updated based solely on the labeled data, allowing us to leverage sparsely annotated videos efficiently. This enables accurate segmentation even when working with partially labeled datasets, maximizing the utility of available annotations while mitigating the impact of unlabeled frames.

\subsection{Evaluation Metrics}
\subsubsection{Spatial and temporal evaluation metrics}
To comprehensively evaluate our segmentation results, we employ a combination of spatial and temporal metrics. For assessing spatial segmentation precision, we utilize three established metrics: the Dice coefficient, which measures overlap between predicted and ground truth; the Hausdorff distance, which quantifies the maximum distance between two point sets; and the Average Symmetric Surface Distance, which provides an overall measure of boundary similarity. 

To complement these spatial assessments, we evaluate temporal consistency using the temporal consistency indicator \cite{painchaud2022echocardiography}. For this temporal evaluation, we first normalize each video to a range of 0 to 1, regardless of its total frame count. We then assess temporal smoothness by analyzing how smoothly the total number of segmentation pixels s$_a$ transitions across frames. This analysis involves examining the second-order derivative of s$_a$. A high derivative value suggests periods of substantial variation, while a lower derivative indicates local smoothness. Given the discrete nature of cardiac time frames, we approximate the second-order derivative numerically as follows:
\begin{align}
\frac{d^2 s_a(t)}{dt^2} \approx s_{a, t+1} + s_{a, t-1} - 2s_{a, t}, 
\end{align}
This approximation operates similarly to a Laplacian filter by evaluating the temporal alignment of three consecutive frames within the cardiac cycle. After computing this operation, we take the absolute values and average them over time, represented as \textit{L}. This approach helps quantify the smooth transition from one segmentation of each frame to the next throughout the cardiac cycle.

\subsubsection{Clinical metrics}
The Left Ventricular Ejection Fraction (LVEF) is a crucial measurement that shows the percentage of blood pumped out of the heart’s primary chamber with each heartbeat. The LVEF is calculated as the ratio between two key metrics: Left Ventricular End-Diastolic Volume (LVEDV) and Left Ventricular End-Systolic Volume (LVESV). Following the clinical guidelines for echocardiography \cite{lang2015recommendations}, we evaluated our method using the Simpson biplane method, a common approach to estimate the volume of the LV of the heart. 
Then, the EF can be calculated as follows:
    \begin{align}
    EF = \left( \frac{EDV - ESV}{EDV} \right) \times 100\%
    \end{align}
    where
         \( EF \) is the ejection fraction,
         \( EDV \) is the ED volume,
         \( ESV \) is the ES volume, respectively.

\subsection{Comparison of state-of-the-art methods }
We conducted an extensive comparison of our method against the state-of-the-art (SOTA) approaches in echocardiography segmentation. Our evaluation includes general-purpose segmentation methods, specialized echocardiography methods, and recent adaptations of vision foundation models. Specifically, we compare with general-purpose segmentation models. U-Net++ \cite{zhou2019unet++}, an advanced variant of the original U-Net architecture, featuring nested and dense skip connections to improve feature propagation. U-Transformer \cite{petit2021u} integrates transformer modules into the U-Net architecture, leveraging self-attention mechanisms to improve feature representation in medical image segmentation. STM \cite{oh2019video} employs a memory network to utilize sptio-temporal information from all past frames. And we include two specialized methods for echocardiography: LUNet \cite{leclerc2020lu}, which sequentially combines localization and segmentation methods for echocardiographic images, and EchoNet-Dynamic \cite{ouyang2020video}, a framework that uses 1D temporal CNN and 2D spatial CNN for LV segmentation. Additionally, we also include several recent adaptations of the SAM for medical imaging: Med-SAM \cite{ma2023segment}, Med-SA \cite{wu2023medical}, SAM-US \cite{lin2023samus}, SAM-Att \cite{zhu2024sam}, SAM3D \cite{bui2023sam3d}, MEMSAM\cite{deng2024memsam}, and MedSAM-2 \cite{zhu2024medical}.

Table \ref{table1} presents comparative results for multi-chamber segmentation, including LV$_{endo}$ and LV$_{epi}$. In the CAMUS dataset evaluation, our method consistently outperforms other methods across all regions of interest (ROIs) in both spatial and temporal metrics. For LV$_{endo}$ segmentation, our method improves the Dice score by 1.20\% and the temporal consistency metric by 0.22 compared to the second-best performing approach, Med-SA. More significantly, in the more challenging task of segmenting LV$_{epi}$, which surrounds LV$_{endo}$ and often presents blurred boundaries, our method outperforms the next best approach by a margin of 1.63\% in Dice score and 0.04 in temporal consistency metrics. These results underscore the effectiveness of our temporal adapter in incorporating temporal information. By utilizing information from both preceding and subsequent frames, MediViSTA more effectively resolves ambiguities present in individual frames, especially for structures with blurred or poorly defined boundaries like LV$_{epi}$. This temporal integration is particularly effective in ensuring consistent and accurate segmentations throughout the cardiac cycle. The visual comparison is presented in Fig. \ref{overlay}.

One of the key advantages of using pretrained SAM is its ability to utilize input prompts for segmentation tasks. In our experiment, we demonstrate both the strengths and limitations of this approach when applied to echocardiography segmentation. When the model uses point prompt input, for example, MEMSAM \cite{deng2024memsam} has shown performance comparable to other state-of-the-art methods for LV$_{endo}$ segmentation. However, its performance decreased significantly when segmenting the LV$_{epi}$, with a Dice score below 30\%. Instead, we utilize bounding box prompts to improve the Dice score by specifying the chambers. For this purpose, we implemented a temporal bounding box, which covers all structures along the time axis at both the ED and the ES, when the left ventricle is at its largest and smallest, respectively. This approach improves the Dice score by 0.20\% for LV$_{endo}$. However, for structures such as LV$_{epi}$ surrounding LV$_{endo}$, even the bounding box prompt failed to achieve satisfactory segmentation results, resulting in a Dice score below 30\%. Given these challenges, we concluded that the pursuit of a fully automatic segmentation approach is a more reasonable choice for echocardiographic segmentation tasks. This approach removes the requirement for manual prompt input, potentially resulting in more consistent segmentation outcomes. Our experiments demonstrate that, while prompt-based methods offer flexibility, they may not always be the optimal solution for addressing complex anatomical structures in echocardiography. 

\begin{table}[t]
\renewcommand{\arraystretch}{1.1}
\caption{Pearson correlation comparison with SOTA methods on multi-center dataset. Best results in \textbf{bold}.}
\setlength{\tabcolsep}{3pt}
\begin{tabularx}{0.48\textwidth}{
    >{\raggedright\arraybackslash}p{2.8cm} |
    >{\centering\arraybackslash}X 
    >{\centering\arraybackslash}X 
    >{\centering\arraybackslash}X
}
\toprule
    Method & LVEDV & LVESV & LVEF \\
    \midrule
    \midrule
    EchoNet \cite{chen2017rethinking} & 0.77 & 0.79 & 0.75 \\
    U-Transformer \cite{petit2021u}   & 0.78 & 0.81 & 0.77 \\
    SAM3D \cite{bui2023sam3d}         & 0.81 & 0.80 & 0.78 \\
    \rowcolor{mygray} MediViSTA                     & \textbf{0.86} & \textbf{0.87} & \textbf{0.84} \\
    \bottomrule
\end{tabularx}
\label{table2}
\end{table}

\subsection{Evaluation of generalization performance} 
Generalization capability is an essential property in the medical image. To evaluate our model's generalization ability, we assessed its zero-shot performance by applying models trained on the CAMUS dataset to the Echonet-dynamic and multi-center internal datasets. As shown in Table \ref{table1}, all methods experienced a performance decrease when segmenting cardiac chambers on unseen datasets. Notably, our method achieved significantly better performance than other methods on both the EchoNet-Dynamic and multi-center datasets, demonstrating its effectiveness and robustness. Compared to the second-best non-SAM-adaptation method, U-transformer \cite{petit2021u}, we achieve a 2.17\% increase in Dice score and a 0.38 improvement in temporal consistency. Compared to the second-best SAM adaptation methods, SAM3D, we still maintain superiority with a 1.55\% higher Dice score and a 0.21 increase in temporal consistency.

In the multi-center dataset, our method achieves even more significant improvements. Compared to the second leading method without SAM adaptation, U-Transformer \cite{petit2021u}, our approach achieves improvements in Dice scores of 3.85\%, 4.82\%, and 3.80\% for LV$_{endo}$, LV$_{epi}$, and LA, respectively, along with increases in temporal consistency of 0.15, 0.06, and 0.02 for each structure. Compared to SAM-based methods, our approach surpasses the second leading method, SAM3D \cite{bui2023sam3d}, in both Dice scores and temporal consistency. Specifically, we achieve improvements of 1.45\%, 3.69\%, and 1.33\% in the Dice scores, and 0.16, 0.04, and 0.08 in temporal consistency for LV$_{endo}$, LV$_{epi}$, and LA, respectively. These consistent improvements across different cardiac structures and datasets demonstrate the robustness and generalizability of our proposed method. Further improvements were observed when applying a bounding box prompt to the model, resulting in a 0. 75\% increase in the Dice score for LV$_{endo}$ and a 0.81\% increase for LA. However, the model achieved less than the 30\% Dice score to delineate the LV$_{epi}$ boundaries.

We further evaluated our method on LVEF measurements, a secondary metric derived from segmentation. We compared our approach to the top performing SOTA methods in each of the three categories: domain-specific methods, general segmentation methods, and SAM-based adaptation methods. For each method, we calculated the Pearson correlation between predicted and ground-truth values of LVEDV, LVESV, and LVEF in all test samples. A higher Pearson correlation indicates that the model’s predictions align closely with real-world clinical indices, demonstrating its reliability for clinical application. Our method demonstrated excellent performance, achieving Pearson correlation coefficients of 0.86 for LVEDV, 0.87 for LVESV, and 0.84 for LVEF, as shown in Table \ref{table2}.

\begin{table}[t]
\renewcommand{\arraystretch}{1.3}
\caption{Effects of temporal-fusion attention. Temporal-fusion, temporal, and spatial attention denoted as $T$-$F$, $T$, $S$.}
\begin{tabularx}{\linewidth}{>{\centering\arraybackslash}X | >{\centering\arraybackslash}X >{\centering\arraybackslash}X}
    \hline
    $T$-$F$ & Dice [\%] $\uparrow$ & $L$ $\downarrow$\\
    \hline
    \hline
    $\circ$ & 87.08 {\scriptsize $\pm$ 7.55} & 0.28 {\scriptsize $\pm$ 0.10} \\
    $\circ$ : \: $T \rightarrow S$ & 87.80 {\scriptsize $\pm$ 6.44} & 0.21 {\scriptsize $\pm$ 0.08} \\
    $\bullet$ : \: $S \rightarrow T\text{-}F$ & 88.10 {\scriptsize $\pm$ 6.08} & 0.09 {\scriptsize $\pm$ 0.04} \\
    \rowcolor{mygray} $\bullet$ : \: $T\text{-}F \rightarrow S$ & \textbf{89.65 {\scriptsize $\pm$ 4.74}} & \textbf{0.06 {\scriptsize $\pm$ 0.04}} \\
    \hline
\end{tabularx}
\label{table3}    
\end{table}

\begin{table}[t]
\renewcommand{\arraystretch}{1.3}
\caption{Comparison of model performance with different kernel.}
\centering
\begin{tabularx}{\linewidth}{>{\centering\arraybackslash}X | >{\centering\arraybackslash}X >{\centering\arraybackslash}X}
    \hline
    Kernel type  & Dice [\%] $\uparrow$ & $L$ $\downarrow$ \\
    \hline
    \hline
    Gaussian, $\sigma = 0.5$ & 88.60 {\scriptsize $\pm$ 5.05} & 0.18 {\scriptsize $\pm$ 0.08} \\
    \rowcolor{mygray} Gaussian, $\sigma = 1.0$ & \textbf{89.65 {\scriptsize $\pm$ 4.74}} & \textbf{0.06 {\scriptsize $\pm$ 0.04}} \\
    Bilateral, $\sigma = 1.0$ & 87.82 {\scriptsize $\pm$ 5.78} & 0.18 {\scriptsize $\pm$ 0.07} \\
    Laplacian, $\sigma = 1.0$ & 86.01 {\scriptsize $\pm$ 6.08} & 0.15 {\scriptsize $\pm$ 0.07} \\
    \hline
\end{tabularx}
\label{table4}
\end{table}

\begin{table*}[!t]
    \centering
    \begin{minipage}[t]{0.35\linewidth} 
        \renewcommand{\arraystretch}{1.2}
            \caption{Comparison of model performance w/ $\bullet$ and w/o $\circ$ the frequency feature fusion module.}
        \centering
        \begin{tabular}{l|c|c|c}
            \hline
            FFM & Transform & Dice [\%] & $L$ $\downarrow$ \\
            \hline   
            \hline
            $\circ$   & - & 85.72 \text{\scriptsize $\pm$ 6.69} & 0.19 \text{\scriptsize $\pm$ 0.09}\\
            $\bullet$ & - & 86.04 \text{\scriptsize $\pm$ 6.48} & 0.10 \text{\scriptsize $\pm$ 0.05}\\
            $\bullet$ & Fourier & 88.14 \text{\scriptsize $\pm$ 6.01} & 0.07 \text{\scriptsize $\pm$ 0.05}\\
            \rowcolor{mygray} $\bullet$ & Wavelet & \textbf{89.65 \text{\scriptsize $\pm$ 4.74}} & \textbf{0.06 \text{\scriptsize $\pm$ 0.04}} \\
            \hline
        \end{tabular}
        \label{table5}
    \end{minipage}
    \hspace{0.01\linewidth} 
    \begin{minipage}[t]{0.28\linewidth} 
        \renewcommand{\arraystretch}{1.3}
        \caption{Comparison of model performance across different ranks.}

        \centering
        \begin{tabular}{l|c|c}
            \hline
            r & Dice [\%] & $L$ $\downarrow$ \\
            \hline   
            \hline   
            4  & 83.50 \text{\scriptsize $\pm$ 8.82} & 0.09 \text{\scriptsize $\pm$ 0.05} \\
            8  & 88.15 \text{\scriptsize $\pm$ 6.24} & 0.07 \text{\scriptsize $\pm$ 0.04} \\
            \rowcolor{mygray} 16 & \textbf{89.65 \text{\scriptsize $\pm$ 4.74}} & \textbf{0.06 \text{\scriptsize $\pm$ 0.04}} \\
            32 & 89.67 \text{\scriptsize $\pm$ 4.70} & 0.06 \text{\scriptsize $\pm$ 0.04} \\
            \hline
        \end{tabular}
        \label{table6}
    \end{minipage}
    \hspace{0.01\linewidth} 
    \begin{minipage}[t]{0.33\linewidth} 
    \centering
    \renewcommand{\arraystretch}{1.3}
    \caption{Comparison of model performance across various temporal adapter designs.}
    \begin{tabular}{l|c|c}
        \hline
        Adapter type  & Dice [\%] $\uparrow$ & $L$ $\downarrow$ \\
        \hline
        \hline
        SAM3D \cite{bui2023sam3d} & 88.60 \text{\scriptsize $\pm$ 5.05} & 0.18 \text{\scriptsize $\pm$ 0.08} \\
        Med-SA \cite{wu2023medical} & 87.82 \text{\scriptsize $\pm$ 5.78} & 0.18 \text{\scriptsize $\pm$ 0.07}\\
        Crossframe \cite{khachatryan2023text2video} & 86.01 \text{\scriptsize $\pm$ 6.08} & 0.15 \text{\scriptsize $\pm$ 0.07}\\
        \rowcolor{mygray} Temporal Fusion & \textbf{89.65 \text{\scriptsize $\pm$ 4.74}} & \textbf{0.06 \text{\scriptsize $\pm$ 0.04}}  \\
        \hline
    \end{tabular}
        \label{table7}
        \end{minipage}
\end{table*}

\section{Ablation Studies}
\label{sec:AS}
In this section, we conducted extensive ablation studies on echocardiography datasets to examine the impact of each component in our proposed SAM fine-tuning strategy. Our analysis focuses on six key aspects: (i) the design and effectiveness of temporal fusion attention, (ii) the effectiveness of the frequency feature fusion module, (iii) the impact of parameter-efficient tuning, (iv) the efficacy of various temporal adapter designs, and (v) the impact of modified mask decoder. We further investigate (vi) the influence of different size of pretrained SAM's backbone. Through these studies, our objective is to validate our design choices and how each component contributes to the effectiveness of the model to address the unique challenges of echocardiography.

\subsection{Effects of Temporal-Fusion Attention}
This ablation study aims to demonstrate the importance of temporal-fusion attention in our model. We compared our full model without temporal-fusion attention, which only uses spatial information. The results show that incorporating temporal-fusion attention improves the Dice score by 2.57\% and temporal smoothness by 0.12. We subsequently conducted a comparison between temporal-fusion attention and temporal attention. The results revealed that substituting temporal fusion attention with temporal attention resulted in a 1.85\% decrease in the Dice score and a 0.15 increase in temporal consistency. We also investigated the optimal sequencing of temporal fusion and spatial attention. As shown in Table \ref{table3}, applying spatial attention before temporal fusion attention led to a performance decline. We observed a decrease of 1.55\% in the Dice score and a 0.03 decline in temporal consistency. These findings emphasize the importance of applying temporal attention before spatial attention in our model architecture, aligning with the TimeSformer \cite{bertasius2021space} for video understanding tasks.

Furthermore, we performed ablation experiments to examine how model performance varies with different types of kernels, including Gaussian, bilateral, and Laplacian kernels, each with a kernel size of 5 and $\sigma$ = 1.0. As shown in Table \ref{table4}, the configuration using the Gaussian kernel achieved the best performance in our final model. Therefore, we adopted this Gaussian kernel configuration in our experiments to optimize the performance of the model.

\subsection{Effects of Frequency Feature Fusion Module}
In this ablation study, we performed the following experiments to validate the effectiveness of the frequency feature fusion module. We first removed the frequency feature fusion module. From Table \ref{table5}, it can be observed that without a frequency feature fusion module to inject spatial frequency information with cross-branch attention, Dice decreased by 2.93\%. We also validated our approach using the Fourier transform instead of the wavelet transform, resulting in a Dice decrease of 1.51\%. Furthermore, using a naive CNN encoder without transforming input information led to a decrease in Dice of 3.61\%.

\subsection{Effects of Parameter-efficient Fine-tuning}
To validate the effectiveness of the parameter-efficient tuning approach, FacT, we performed comparisons without and with parameter-efficient fine-tuning, adjusting only a small subset of weight increases. Without the FacT approach, updating only the adapters without fine-tuning SAM leads to reduced accuracies of 89.65\% and 87.02\%, respectively. We also examine how the performance of the model varies with different decomposition ranks, r, selecting values from the set \{4, 8, 16, 32\}. As shown in Table \ref{table6}, increasing the rank generally improves the average performance of Dice, although the gains are saturated around r = 16. To balance performance with additional parameters, we set r = 16 in our experiments.

\begin{table}[t]
\renewcommand{\arraystretch}{1.3}
\caption{An ablation study on the effect of multi-scale fusion on modified mask decoder design.}
\begin{tabularx}{\linewidth}{>{\centering\arraybackslash}X | >{\centering\arraybackslash}X >{\centering\arraybackslash}X}
    \hline
    Multi-scale fusion     & Dice [\%] $\uparrow$ & $L$ $\downarrow$ \\
    \hline
    \hline
    $\circ$ & 86.12 \text{\scriptsize $\pm$ 7.10} & 0.31 \text{\scriptsize $\pm$ 0.23} \\
    \rowcolor{mygray} $\bullet$ & \textbf{89.65  \text{\scriptsize $\pm$ 4.74}} & \textbf{0.06 \text{\scriptsize $\pm$ 0.04}} \\
    \hline
\end{tabularx}
\label{table8}
\end{table}

\subsection{Impact of Various Temporal Adapter Types}
In Table \ref{table7}, we explore alternative methods for incorporating temporal information to further validate our proposed approach. We tested several alternatives: a 3D CNN encoder similar to \cite{bui2023sam3d}, an adapter that divides attention into spatial and temporal branches as in \cite{wu2023medical}, and a cross-frame attention mechanism \cite{khachatryan2023text2video}. The 3D CNN approach decreased the Dice score by 1.05\% and temporal consistency by 0.08. The bifurcated attention adapter led to a 1.83\% decrease in the Dice score and a 0.08 reduction in temporal consistency. We also compared our temporal-fusion attention to a similar mechanism that utilizes cross-frame attention \cite{khachatryan2023text2video}. The cross-frame approach calculates $Q$ on the current frame and derives $K$ and $V$ from the first frame to constrain the generated video throughout time. However, this method can negatively affect video segmentation if the first frame is not visible or is of poor quality. Our experiments showed that the approach of cross-frame attention led to a decrease of 3.64\% in the Dice score and an increase of 0.09 in temporal consistency compared to our method. In contrast, our temporal-fusion attention utilizes adjacent frames, making it more effective and robust, especially in scenarios where the quality or visibility of the previous frame may be compromised.

\subsection{Effects of Modified Mask Decoder}
This study evaluates the effectiveness of our modifications to mask decoder components. As presented in Table \ref{table8}, we observed a 3.53\% increase in Dice score and a 0.25 improvement in temporal consistency. These results highlight the importance of improving prediction resolution in echocardiography, particularly given the variable sizes of chambers that can constitute regions of interest. Our multiscale fusion approach effectively addresses the challenges posed by the diverse spatial characteristics of echocardiographic data, contributing to more robust and accurate segmentation results in varying chamber sizes.

\subsection{Effects of SAM's Backbone}
This study compares the performance of pretrained SAM with different network backbones: ViT-B, ViT-L, and ViT-H. As observed in Table \ref{table9}, the largest pretrained model, ViT-H, consistently demonstrates superior accuracy in all metrics, achieving the highest Dice score and exhibiting the best temporal consistency with 89.65 and 0.06, respectively. However, performance decreases as the size of the model decreases, with ViT-B showing the lowest Dice score of 87.98\%. These results highlight that increasing model size enhances overall performance.

\begin{table}[t]
\renewcommand{\arraystretch}{1.3}
\caption{An ablation study of model performance using different sizes of pretrained SAM backbones.}
\begin{tabularx}{\linewidth}{>{\centering\arraybackslash}X | >{\centering\arraybackslash}X >{\centering\arraybackslash}X}
    \hline
    SAM's Backbone      & Dice [\%] $\uparrow$ & $L$ $\downarrow$ \\
    \hline
    \hline
    ViT-B & 87.98 \text{\scriptsize $\pm$ 6.04} & 0.26 \text{\scriptsize $\pm$ 0.05} \\
    ViT-L & 88.68 \text{\scriptsize $\pm$ 4.72} & 0.15 \text{\scriptsize $\pm$ 0.04} \\
    \rowcolor{mygray}ViT-H & \textbf{89.65  \text{\scriptsize $\pm$ 4.74}} & \textbf{0.06 \text{\scriptsize $\pm$ 0.04}}\\
    \hline
\end{tabularx}
\label{table9}
\end{table}

\section{Discussion}
We propose a new parameter-efficient SAM adaptation method to adapt SAM from a 2D natural image to medical video imaging, especially for echocardiography segmentation. The modifications to the image encoders are specifically designed to accommodate video input while leveraging pretrained weights. For both spatial and temporal adaptation, we introduce a frequency feature fusion module and a temporal-fusion attention mechanism. The mask decoder utilizes multiscale aggregation to leverage intermediate features more effectively from the encoder. Our model uses a subset of parameters, comprising 1.9\% of the learnable parameters, with a computational cost of 45.41 GFLOPs and an inference speed of 1.56 FPS. Experiments on public and multicenter echocardiography segmentation datasets demonstrate the superiority of our approach over state-of-the-art medical image segmentation models and current parameter-efficient fine-tuning methods.

One significant motivation for adapting general purpose vision foundation models for medical images is its pretraining on a massive and diverse dataset, which is difficult to achieve in the field of medical imaging. General-purpose foundation models hold promise for significant advancements in medical applications. As shown in Table \ref{table1}, pre-training in general-purpose data benefits medical image segmentation. Recently, memory-based models like SAM2 and MEMSAM have extended 2D SAM for video segmentation by incorporating streaming memory, which stores previous prompts and predictions. However, these models often face limitations due to their dependency on prior frames. For instance, they rely heavily on conditions set by preceding frames, which can limit their flexibility and accuracy when segmenting the current frames. Our temporal fusion attention method, which uses kernels to weight adjacent frames during attention, has proven efficient for echocardiography segmentation as in Table \ref{table3}.

Although MediViSTA achieves notable results with unseen datasets, there is still room for further refinement. The framework is designed to work with pre-acquired and saved image scans, focusing its capabilities on detailed post-acquisition analysis. This design choice has yet to be extended to real-time processing during image acquisition, which is often essential for echo scans. This feature could be particularly valuable in point-of-care settings, where rapid interpretation is essential, especially for personnel who may lack specialized expertise in scanning echocardiography images.

Future research will focus on building domain-specific foundation models tailored for echocardiography. As medical datasets continue to accumulate in databases, we will leverage these data resources to build a foundation model from scratch and improve model accuracy, robustness, and generalizability, ultimately improving diagnostic and clinical outcomes in echocardiography. In addition, we will address our limitations and expand our capabilities, focusing particularly on real-time processing and model efficiency \cite{zhao2024no}, to enhance its utility as a point-of-care tool for non-expert operators. This would involve optimizing the model's architecture and inference speed to process echocardiographic data streams in real-time without compromising accuracy. These optimizations could enable deployment on portable ultrasound machines, making advanced echocardiographic analysis more accessible in resource-limited settings. 

\section{Conclusion}
In this study, we propose a comprehensive approach to adapting the Segment Anything Model (SAM) from 2D natural images to medical video segmentation, with a focus on echocardiography. By applying parameter-efficient fine-tuning and extending the SAM to incorporate third-dimensional information, our method significantly improves the performance of the SAM in the medical domain, outperforming current state-of-the-art methods. In particular, our approach demonstrates strong results in zero-shot analysis across multi-center datasets, underscoring its potential for generalization in diverse clinical settings. In particular, our approach demonstrates strong results in zero-shot analysis in multi-center data sets, underscoring its potential for generalization in diverse clinical settings. The implementation is open-source at: \href{https://github.com/SekeunKim/MediViSTA.git}{https://github.com/SekeunKim/MediViSTA.git}, to support further research and development in echocardiography.

\bibliographystyle{IEEEtran} 
\bibliography{ref}

\end{document}